\documentclass{emulateapj}

\shorttitle{First measurements of the electron density in C--shocks}
\shortauthors{I. Jim\'{e}nez--Serra et al.}

\usepackage{multirow}
\begin{document}

\title{First Measurements of the electron density enhancement expected in C--shocks.}

%\title{Ion and Neutral Excitation Differentiation in the L1448--mm Outflow}

\author{I. Jim\'{e}nez--Serra\altaffilmark{1}, J.
  Mart\'{\i}n--Pintado\altaffilmark{1}, S. Viti\altaffilmark{2}, 
  S. Mart\'{\i}n\altaffilmark{3}, A.
  Rodr\'{\i}guez--Franco\altaffilmark{1,4}, 
  A. Faure\altaffilmark{5} and
  J. Tennyson\altaffilmark{2}}

\altaffiltext{1}{Departamento de Astrof\'{\i}sica Molecular e Infrarroja, 
Instituto de Estructura de la Materia (CSIC),
C/ Serrano 121, E--28006 Madrid, Spain; 
izaskun@damir.iem.csic.es, martin@damir.iem.csic.es,
arturo@damir.iem.csic.es}

\altaffiltext{2}{Department of Physics and Astronomy, University
  College London, London WC1E$\,$6BT, England, UK; sv@star.ucl.ac.uk,
  j.tennyson@ucl.ac.uk}

\altaffiltext{3}{Instituto de Radioastronom\'{\i}a Milim\'etrica,
Avda. Divina Pastora, Local 20, E--18012 Granada, Spain; martin@iram.es}

\altaffiltext{4}{Escuela Universitaria de \'Optica,  
Departamento de Matem\'atica Aplicada (Biomatem\'atica),
Universidad Complutense de Madrid,
Avda. Arcos de Jal\'on s/n, E--28037 Madrid, Spain}

\altaffiltext{5}{Laboratoire d'Astrophysique, Observatoire de
  Grenoble, BP$\,$53, 38041 Grenoble Cedex$\,$09, France;
  Alexandre.Faure@obs.ujf-grenoble.fr} 

\begin{abstract}

Magnetic precursors of C--shocks accelerate, compress
and heat molecular ions, modifying the kinematics and the 
physical conditions of the ion fluid with respect to the
neutral one. Electron densities are also expected to be significantly 
enhanced in shock precursors. In this Letter,
we present observations of strongly polar ion and neutral
molecules such as SiO, H$^{13}$CO$^{+}$, HN$^{13}$C and H$^{13}$CN, which
reveal the electron density enhancements associated 
with the precursor of the young L1448--mm outflow. While
in the ambient gas the excitation of the ions and neutrals 
is explained by collisional excitation by H$_2$ with a single 
density of $\sim$10$^5$$\,$cm$^{-3}$, H$^{13}$CO$^{+}$ shows an over
excitation in the shock precursor component that requires H$_2$ densities of
a factor of $\geq$10 larger than those derived from the neutral
species. This over excitation in H$^{13}$CO$^{+}$ can be explained if
we consider an additional excitation by collisions with electrons and
an electron density enhancement in the precursor stage by a factor of 
$\sim$500, i.e. a fractional ionization of 5$\times$10$^{-5}$.
These results show that 
multiline observations can be used to study the evolution of 
the ion and electron fluids at the first stages of the
C--shock interaction.

\end{abstract}


\section{Introduction}

In dense molecular clouds, the fractional ionization of the gas 
is very low \citep[$\leq$10$^{-7}$; Gu\'elin, Langer \& Wilson
1982;][]{cas98}. Electron densities are therefore 
insufficient to excite the first rotational 
levels of even strongly polar molecules like HCO$^{+}$, 
HCN or HNC %\citep[$\mu$$\sim$3--4$\,$Debyes;][]{sto66,hae79} 
whose collisional cross sections
\citep[$\sim$10$^{-6}$--10$^{-5}$$\,$cm$^3$$\,$s$^{-1}$; 
Bhattacharyya, Bhattacharyya \& Narayan 1981;][]{saha81,fau01}
are a factor of $\geq$100 larger than those of weakly polar 
species like CO
\citep[$\sim$10$^{-8}$$\,$cm$^3$$\,$s$^{-1}$;][]{saha81}.  
%and $\mu$$\sim$0.11$\,$Debyes;][]{sto66,saha81}. 
All attempts to use molecular excitation
to measure the effects of collisional excitation by electrons 
in these regions have been so far unsuccessful \citep{lan85}.

Modeling of the early stages of C--shocks predicts that the ion and electron
densities are enhanced by the magnetic precursor \citep{dra80}.
For molecules like HCO$^{+}$, HCN or HNC,
the electron density enhancement by a factor of $\sim$100 produced
by the precursor \citep[see e.g.][]{flo96},
would make electron collisions competitive 
with excitation by H$_2$ collisions for the typical densities of dark clouds
($\sim$10$^5$$\,$cm$^{-3}$). Since the electron collisional 
coefficients for high initial \textit{J} transitions of HCO$^{+}$ are 
a factor of $\geq$10 larger than those of neutral molecules like HCN 
at low temperatures \citep[$\sim$10--20$\,$K;][]{chu74,bha81}, 
differences in the molecular excitation between ion  
and neutral species are therefore expected in the precursor stage 
\footnote{These excitation differences between ion and neutral molecules
reflect a fundamental difference between the corresponding electron--impact
cross sections at low energy: for neutrals, cross sections go to zero at
threshold while for ions, they are large and finite
\citep[e.g.][]{chu74}.}.
Toward the young L1448--mm molecular outflow, the
detection of very narrow SiO emission and the enhancement of the ion
abundance have been interpreted as signatures of the shock precursor
\citep{jim04}. Multiline observations of strongly polar species
like HCO$^{+}$, HNC and HCN toward this outflow, 
are expected to show differences in
excitation between the ambient gas where electron excitation is
negligible, and the shock precursor where electron collisions should
be important.   

In this Letter, we present observations of several rotational transitions of 
SiO, H$^{13}$CO$^{+}$, HN$^{13}$C and H$^{13}$CN
observed toward the ambient and shock precursor components 
of the L1448--mm outflow. 
The excitation differences observed between H$^{13}$CO$^{+}$, and SiO,
HN$^{13}$C and H$^{13}$CN in the ambient and precursor components, 
can be explained by the electron density enhancement expected 
at the first stages of the C--shock evolution.

%FIG1**************************************
\begin{figure}
\epsscale{1.05}
\plotone{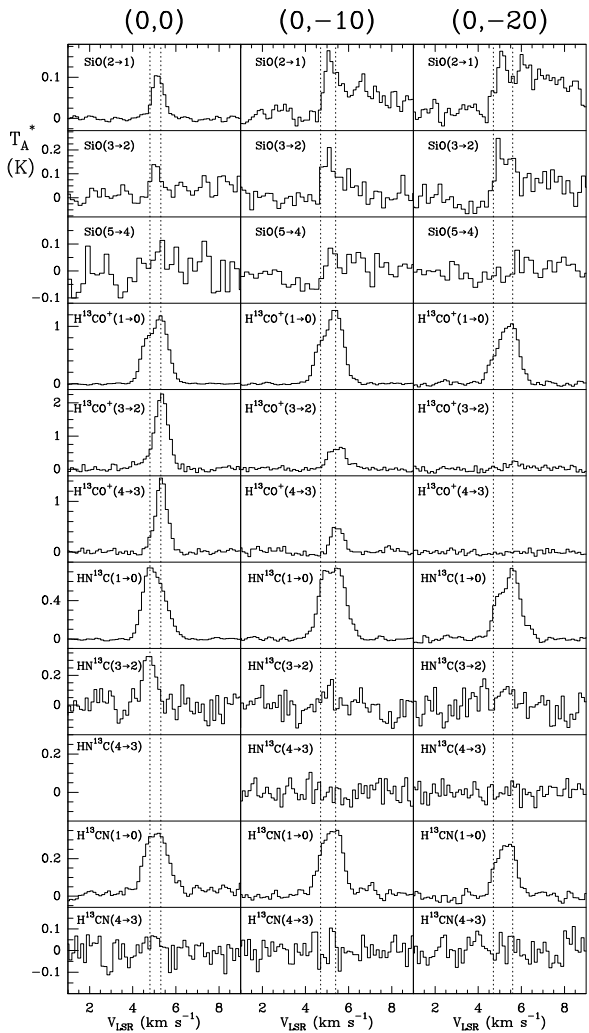}
\caption{Low-- and high--\textit{J} lines of SiO,
  H$^{13}$CO$^{+}$, HN$^{13}$C and H$^{13}$CN 
  observed toward L1448--mm. Offsets in
  arcseconds are shown in the upper part of the columns 
  ($\alpha_{(J2000)}$~=~03$^{h}$25$^{m}$38$_.^s$0, 
  $\delta_{(J2000)}$~=~30$^{\circ}$44$'$05$''$). 
  The vertical dotted lines show the ambient gas at 4.7$\,$km$\,$s$^{-1}$ 
  and the precursor component at 5.2$\,$km$\,$s$^{-1}$ toward
  L1448--mm (0,0), and at 5.4 and 5.6$\,$km$\,$s$^{-1}$ toward
  L1448--mm (0,--10) and (0,--20), respectively.}
\label{fig1}
\end{figure}
%******************************************

\section{Observations \& Results}

We observed several transitions (from \textit{J}$\,$=1 to 5)
of SiO, H$^{13}$CO$^{+}$, HN$^{13}$C and
H$^{13}$CN toward three different positions in the L1448--mm outflow. 
Except the \textit{J}$\,$=$\,$4$\rightarrow$3 lines of
H$^{13}$CO$^{+}$, HN$^{13}$C and H$^{13}$CN, all the molecular
transitions were observed with the
IRAM 30$\,$m telescope at Pico Veleta (Spain). We used the
wobbler--switched and frequency--switched modes with position 
and frequency throws of 240$''$ and 7.2$\,$MHz. The beam size was
$\sim$28$''$,  19$''$ and 11$''$ at $\sim$90, 130 and 260$\,$GHz.  
The SIS receivers were tuned to single--side band with rejections of
$\geq$10$\,$dB. We used the VESPA spectrometers with a spectral
resolution of 40$\,$kHz, i.e. a velocity resolution of $\sim$0.14, 0.09 and
0.05$\,$km$\,$s$^{-1}$ at $\sim$90, 130 and 260$\,$GHz. Typical system
temperatures ranged from 100 to 500$\,$K. 

The \textit{J}$\,$=$\,$4$\rightarrow$3 lines ($\sim$347$\,$GHz) of
H$^{13}$CO$^{+}$, HN$^{13}$C and H$^{13}$CN 
were observed with the 
JCMT telescope at Mauna Kea (Hawaii) in the frequency--switched mode 
with a frequency throw of 16$\,$MHz. The beam size was
$\sim$14$''$, which matches the IRAM 30$\,$m beam for the
\textit{J}$\,$=3$\rightarrow$2 lines of H$^{13}$CO$^{+}$ and
HN$^{13}$C. We used the B3 receiver in dual--mixer 
and single--side band mode with an image rejection of 12--14$\,$dB. 
The DAS spectrometer provided a spectral resolution of 156$\,$KHz
($\sim$0.14$\,$km$\,$s$^{-1}$). The system temperatures
were of 450--560$\,$K. All the intensities were calibrated
in units of antenna temperature and converted to main beam
temperatures using efficiencies of 0.82, 0.74 and 0.52 at $\sim$90, 130 
and 260$\,$GHz for the IRAM 30$\,$m data, and 0.63 for the JCMT data.

Fig.$\,$1 shows the line profiles of all transitions 
measured toward L1448--mm (0,0), (0,--10) and (0,--20), and 
Tab.$\,$1 gives the observed parameters for the different velocity components.
%The line profiles of SiO, the ion (H$^{13}$CO$^{+}$) 
%and the neutral (HN$^{13}$C and H$^{13}$CN) species 
%show appreciable differences. 
As expected for a shock tracer in the
precursor (Jim\'enez--Serra et al. 2004), 
the narrow ($\sim$0.6$\,$km$\,$s$^{-1}$) SiO 
\textit{J}$\,$=$\,$2$\rightarrow$1 and 3$\rightarrow$2 
lines have single--peaked profiles whose
peak emission is slightly redshifted ($\sim$5.2$\,$km$\,$s$^{-1}$; Fig.$\,$1)
with respect to the ambient 4.7$\,$km$\,$s$^{-1}$ cloud. The
\textit{J}$\,$=$\,$1$\rightarrow$0 lines of
H$^{13}$CO$^{+}$, HN$^{13}$C and H$^{13}$CN show 
double--gaussian profiles 
%with velocity peaks at 4.7$\,$km$\,$s$^{-1}$ and 5.2$\,$km$\,$s$^{-1}$ 
(the ambient and shock precursor components)
and have linewidths of $\sim$0.6--0.7$\,$km$\,$s$^{-1}$ for each of the
velocity components. The H$^{13}$CO$^{+}$ emission 
peaks at the shock precursor component toward the positions where 
narrow SiO is detected (Jim\'enez--Serra et al. 2004). However, 
the HN$^{13}$C peak emission is centered at the
ambient cloud in L1448--mm (0,0). Toward L1448--mm (0,--10) and (0,--20), 
the HN$^{13}$C emission is progressively redshifted 
to 5.4 and 5.6$\,$km$\,$s$^{-1}$, respectively 
(see vertical doted lines in Fig.$\,$1). 
H$^{13}$CO$^{+}$ and H$^{13}$CN also peak at 5.4 and 5.6$\,$km$\,$s$^{-1}$
toward these positions, as if we were observing the evolutionary
effects of the propagation of C--shocks through the unperturbed
gas. The detection of broad SiO emission (terminal velocity of
$\sim$25$\,$km$\,$s$^{-1}$) toward 
L1448--mm (0,--10) and (0,--20), supports this idea. 

%In contrast with the \textit{J}$\,$=$\,$1$\rightarrow$0 transition
%of H$^{13}$CO$^{+}$, HN$^{13}$C and H$^{13}$CN, 
The \textit{J}$\,$=$\,$3$\rightarrow$2 and
4$\rightarrow$3 lines of H$^{13}$CO$^{+}$, HN$^{13}$C and H$^{13}$CN 
tend to be single--peaked and have 
linewidths of $\sim$0.7$\,$km$\,$s$^{-1}$ (Tab.$\,$1). The 
HN$^{13}$C emission arising from the ambient gas toward L1448--mm (0,0)
is weak. 
%and is only detected toward L1448--mm (0,0).
However, the high--\textit{J} H$^{13}$CO$^{+}$ emission measured
toward L1448--mm (0,0) and (0,--10), is very bright and
mainly arises from the precursor component. 
H$^{13}$CO$^{+}$ also shows faint emission
centered at the ambient cloud toward L1448--mm (0,0). 
The SiO \textit{J}$\,$=$\,$5$\rightarrow$4 and the H$^{13}$CN
\textit{J}$\,$=$\,$4$\rightarrow$3 lines have not been detected
in any of the velocity components toward L1448--mm.

\section{Excitation Differences between Ion and Neutral Molecular Species}

%TABLE 1*****************************************************************

\begin{deluxetable*}{lccccccccc}
\tabletypesize{\scriptsize}
\tablecaption{Observed parameters of SiO, ion and neutral species toward
  L1448--mm. \label{tbl-1}}
\tablewidth{0pt}
\startdata
& & & \multicolumn{3}{c}{TABLE 1} & & & & \\
& & \bf{(0,0)} & & & \bf{(0,--10)} & & & \bf{(0,--20)} & \\
\hline \hline
Line & V$_{LSR}$ & $\Delta$$v$ & T$_A$$^*$ & V$_{LSR}$ &
$\Delta$$v$ & T$_A$$^*$ & V$_{LSR}$ & $\Delta$$v$ & T$_A$$^*$ \\
 & (km$\,$s$^{-1}$) & (km$\,$s$^{-1}$) & (K) & (km$\,$s$^{-1}$) &
(km$\,$s$^{-1}$) & (K) & (km$\,$s$^{-1}$) & (km$\,$s$^{-1}$) & (K) \\ \hline

SiO(2$\rightarrow$1) & $\sim$4.7 & $\ldots$ & $\leq$0.012 &
$\sim$4.7 & $\ldots$ & $\leq$0.048 & $\sim$4.7 & $\ldots$ &
$\leq$0.060 \\
& 5.170(8) & 0.62(2) & 0.107(5) & 5.18(3) & 0.60(8)
& 0.11(2) & 5.11(2) & 0.41(8) & 0.10(2) \\

SiO(3$\rightarrow$2) & $\sim$4.7 & $\ldots$ & $\leq$0.084 &
$\sim$4.7 & $\ldots$ & $\leq$0.132 & $\sim$4.7 & $\ldots$ &
$\leq$0.183 \\
& 5.08(7) & 0.6(2) & 0.14(1) & 5.02(6) & 0.5(1) & 0.17(3) & 5.20(7) &
0.9(1) & 0.18(4) \\

SiO(5$\rightarrow$4) & $\sim$4.7 & $\ldots$ & $\leq$0.171 & $\sim$4.7 & $\ldots$ & $\leq$0.130 &
$\sim$4.7 & $\ldots$ & $\leq$0.084 \\
& $\sim$5.2 & $\ldots$ & $\leq$0.171 &
$\sim$5.2 & $\ldots$ & $\leq$0.130 & $\sim$5.2 & $\ldots$ &
$\leq$0.084 \\

H$^{13}$CO$^{+}$(1$\rightarrow$0) & 4.584(7) & 0.59(1) &
0.629(8) & 4.685(6) & 0.70(1) & 0.62(2) & 4.84(2) &
0.84(5) & 0.46(3) \\
& 5.300(5) & 0.81(1) & 1.160(8) & 5.424(0) & 0.796(8) &
1.26(2) & 5.545(9) & 0.84(2) & 1.00(3) \\

H$^{13}$CO$^{+}$(3$\rightarrow$2) & 4.7(0) & 0.6(0) &
0.28(8) & $\sim$4.7 & $\ldots$ & $\leq$0.168 &
$\sim$4.7 & $\ldots$ & $\leq$0.225 \\
& 5.356(6) & 0.69(2) & 2.25(8) & 5.48(2) & 0.95(5) &
0.66(8) & $\sim$5.2 & $\ldots$ & $\leq$0.225 \\

H$^{13}$CO$^{+}$(4$\rightarrow$3) & 4.7(0) & 0.55(0)
& 0.08(5) & $\sim$4.7 & $\ldots$ & $\leq$0.162 & $\sim$4.7 &
$\ldots$ & $\leq$0.138 \\ 
& 5.338(7) & 0.62(2) & 1.43(5) & 5.47(2) & 0.63(4) &
0.52(4) & $\sim$5.2 & $\ldots$ & $\leq$0.138 \\

HN$^{13}$C(1$\rightarrow$0) & 4.785(3) & 0.806(7) & 0.72(1)
& 4.81(2) & 0.65(2) & 0.56(2) & 4.85(2) & 0.50(4) & 0.34(3) \\
& 5.468(6) & 0.78(2) & 0.36(1) & 5.53(2) & 0.88(3) &
0.71(2) & 5.59(1) & 0.86(4) & 0.68(3) \\

HN$^{13}$C(3$\rightarrow$2) & 4.75(5) & 0.6(1) & 0.35(4) & $\sim$4.7 &
$\ldots$ & $\leq$0.261 & $\sim$4.7 & $\ldots$ & $\leq$0.264 \\
& $\sim$5.2 & $\ldots$ & $\leq$0.237 & $\sim$5.2 & $\ldots$ &
$\leq$0.261 & $\sim$5.2 & $\ldots$ & $\leq$0.264 \\ 

HN$^{13}$C(4$\rightarrow$3) & 4.38(8) & 0.8(0) & 0.16(3) & $\sim$4.7 &
$\ldots$ & $\leq$0.117 & $\sim$4.7 & $\ldots$ & $\leq$0.108 \\
& $\sim$5.2 & $\ldots$ & $\leq$0.144 & $\sim$5.2 & $\ldots$ &
$\leq$0.117 & $\sim$5.2 & $\ldots$ & $\leq$0.108 \\ 

H$^{13}$CN(1$\rightarrow$0) & 4.86(5) & 1.06(7) & 0.18(2) & 4.89(2) &
0.80(6) & 0.27(1) & 4.9(1) & 0.6(2) & 0.16(1) \\ 
& 5.32(2) & 1.25(4) & 0.21(2) & 5.52(2) & 0.70(6) & 0.30(1) & 5.5(1) &
0.7(2) & 0.28(1) \\

H$^{13}$CN(4$\rightarrow$3) & $\sim$4.7 & $\ldots$ & $\leq$0.141 &
$\sim$4.7 & $\ldots$ & $\leq$0.138 & $\sim$4.7 & $\ldots$ & $\leq$0.141 \\
& $\sim$5.2 & $\ldots$ & $\leq$0.141 & $\sim$5.2 & $\ldots$ &
$\leq$0.138 & $\sim$5.2 & $\ldots$ & $\leq$0.141\\ \hline
\enddata
\end{deluxetable*}
%***************************************************************************

A first look to the high--\textit{J} lines of H$^{13}$CO$^{+}$,
HN$^{13}$C and H$^{13}$CN, clearly shows that 
the emission of H$^{13}$CO$^{+}$ is anomalously bright 
in the shock precursor component compared to that of the neutrals. 
Since the emission of all these species is expected to be 
optically thin, the line intensity
ratio between different transitions is directly related to the excitation 
temperature of the ion and neutral fluids in the ambient and shock precursor
gas. The integrated line ratios between the
\textit{J}$\,$=$\,$3$\rightarrow$2 and 1$\rightarrow$0 lines for
H$^{13}$CO$^{+}$ and HN$^{13}$C, and between the 
\textit{J}$\,$=$\,$4$\rightarrow$3 and 1$\rightarrow$0 lines for
H$^{13}$CO$^{+}$ and H$^{13}$CN in the ambient and precursor components, 
are shown in Tab.$\,$2. In the ambient cloud, the line ratios 
of H$^{13}$CO$^{+}$ are very similar to those of HN$^{13}$C and
H$^{13}$CN. However, the line intensity ratios of
H$^{13}$CO$^{+}$ in the precursor component are up to a factor of
9 larger than those of the neutrals, indicating a higher excitation
for the ions in this component. H$^{13}$CO$^{+}$ is 
``selectively'' excited in the precursor gas. 

We can estimate the H$_2$ densities required to explain the 
line intensity ratios of Tab.$\,$2, by using a model for the excitation 
of the observed molecules. Let us consider the LVG approximation and
the only excitation by H$_2$ collisions. We have used
the H$_2$ collisional rates of \citet{tur92} for SiO, \citet{flo99}
for H$^{13}$CO$^{+}$, and \citet{gre74} for
HN$^{13}$C and H$^{13}$CN. Considering that the emissions of all
molecules have similar spatial distributions, and a kinetic
temperature of 21$\,$K \citep[see][]{cur99},
the estimated H$_2$ densities and molecular column densities
are shown in Tab.$\,$3. For the ambient gas, 
the derived H$_2$ densities for all molecules are of few
10$^{5}$$\,$cm$^{-3}$, consistent with excitation by H$_2$
collisions. The H$_2$ densities derived from SiO, HN$^{13}$C and 
H$^{13}$CN in the precursor are similar to those of the
ambient gas.
%The H$_2$ densities required to explain the 
%line intensities of SiO, HN$^{13}$C and
%H$^{13}$CN in the shock precursor component, 
%are similar to those of the ambient gas.
However, as expected from the large H$^{13}$CO$^{+}$ line ratios, 
the H$_2$ densities required for this ion in the precursor gas 
are a factor of $\geq$10 larger than those for 
the neutral species (Tab.$\,$3). This clearly illustrates that
excitation only by H$_2$ collisions with a single density 
cannot explain the excitation of H$^{13}$CO$^{+}$ in the precursor.

%TABLE 2***************************************************************
\begin{deluxetable}{llcccc}
\tabletypesize{\scriptsize}
%\rotate
\tablecaption{Integrated line intensity ratios \label{tbl-1}}
\tablewidth{0pt}
\startdata
%\hline \hline
& & \multicolumn{2}{c}{TABLE 2} & & \\
& &
\multicolumn{2}{c}{\bf{(0,0)}} & \multicolumn{2}{c}{\bf{(0,--10)}} \\
\hline \hline
%\cline{2-6}
& & {\bf Amb.} & {\bf Pre.} & {\bf Amb.} & {\bf Pre.} \\
\multirow{2}{*}{{\large$\left(\frac{3\rightarrow2}{1\rightarrow0}\right)$}}
& H$^{13}$CO$^{+}$ & 0.5 & 1.6 & $\leq$0.1 & 0.6 \\
& HN$^{13}$C & 0.4 & $\leq$0.2 & $\leq$0.2 & $\leq$0.1 \\
\cline{2-6}
\multirow{2}{*}{{\large$\left(\frac{4\rightarrow3}{1\rightarrow0}\right)$}}
& H$^{13}$CO$^{+}$ & 0.1 & 0.9 & $\leq$0.1 & 0.3 \\
& H$^{13}$CN & $\leq$0.2 & $\leq$0.1 & $\leq$0.2 & $\leq$0.2 \\ \hline

\enddata
\end{deluxetable}
%********************************************************************

\section{Collisional Excitation by Electrons.}

Since the ions have been selectively excited in the precursor
component by an extra mechanism beside the H$_2$ impact, 
we explore the possibility that this selective excitation 
is produced by collisions with electrons.
%From the H$_2$ densities derived for the precursor 
%component in section$\,$3, it seems that the ions
%have been selectively excited in this component by an extra mechanism 
%beside the H$_2$ impact. In the following, we explore the 
%possibility that this selective excitation is produced by
%collisions with electrons. 
The efficiency of excitation of molecular 
ions by electrons can be significantly larger than that of
neutral molecules at the low temperatures of dark clouds. To
illustrate this, we compare the electron collisional rates of
HCO$^{+}$ and HCN for the \textit{J}$\,$=$\,$0$\rightarrow$1 
and 1$\rightarrow$2 transitions at low and high temperatures. 
While the HCO$^{+}$/HCN collisional coefficient ratio 
is only $\sim$1.6 at 100$\,$K, this ratio is increased to $\sim$6 at
a temperature of 10$\,$K \citep{bha81,saha81}.
This difference between the HCO$^{+}$ and HCN rates is expected 
to further increase for higher initial \textit{J}
and large $\Delta$\textit{J} transitions \citep{bha81}. New 
calculations of the HCO$^{+}$ and H$^{13}$CO$^{+}$
collisional rates for all transitions between \textit{J}$\,$=$\,$1 and 
\textit{J}$\,$=$\,$5 \citep{fau06},
show that the HCO$^{+}$ rates with $\Delta$\textit{J}$\geq$2 
exceed those of HCN by more than one order of magnitude 
at 10$\,$K. %\citep[see also][]{saha81}. 
This naturally introduce a differential excitation 
between the ions and neutrals as observed in the precursor component.

We can constrain the electron density required to reproduce
the H$^{13}$CO$^{+}$ line intensities observed in the precursor component 
by using the LVG model including collisions with both H$_2$ and electrons. 
For an H$_2$ density of $\sim$3$\times$10$^5$$\,$cm$^{-3}$ 
(similar to that derived from
HN$^{13}$C and H$^{13}$CN; Tab$\,$3) and a temperature of 21$\,$K, 
the estimated electron densities in the precursor component
toward L1448--mm (0,0) and (0,--10) in the optically thin case
[N(H$^{13}$CO$^{+}$)$\sim$10$^{11}$$\,$cm$^{-2}$], are $\sim$240 and
600$\,$cm$^{-3}$ respectively, which correspond to fractional
ionizations of 8$\times$10$^{-4}$ and 2$\times$10$^{-3}$. 
However, in the optically thick case 
[N(H$^{13}$CO$^{+}$)$\sim$7--9$\times$10$^{12}$$\,$cm$^{-2}$], and
for higher temperatures ($\sim$35--45$\,$K), the fractional
ionization in the precursor decreases to
$\sim$5$\times$10$^{-5}$. Although even higher temperatures
($\sim$100$\,$K) could reproduce the H$^{13}$CO$^{+}$
intensities in the precursor component for a fractional
ionization of $\leq$10$^{-7}$, the derived HN$^{13}$C and
H$^{13}$CN line intensities would clearly exceed (by up to a factor of 5) 
the upper limits of Tab.$\,$1. The derived ionization fraction implies an 
electron density enhancement by a factor of $\sim$500 with respect to that of 
the quiescent gas ($\leq$10$^{-7}$).

Considering an extrapolation of the electron collisional rates 
of \citet{saha81} for HCN, we can now estimate the expected line 
intensities of HN$^{13}$C and H$^{13}$CN in the precursor for the fractional
ionization ($\sim$5$\times$10$^{-5}$) derived from H$^{13}$CO$^{+}$.
The expected intensities are similar to those reported in
Tab.$\,$1, except for the HN$^{13}$C
\textit{J}$\,$=$\,$3$\rightarrow$2 line whose predicted intensity 
exceeds the upper
limits in Tab.$\,$1 by a factor of $\sim$2. Given the uncertainties in
the rates, and the relative spatial distribution
of the ion and neutral gas, the data are
consistent with the idea of an electron density enhancement in the
precursor. High angular resolution observations are required
to establish the spatial distribution of the ion and neutral species
in this component.

\section{On the Origin of the Electron Density Enhancement}

Toward the quiescent gas of L1448--mm, the ion and neutral 
fluids show similar excitation conditions. In fact, 
the H$_2$ densities obtained from the ions and neutrals are all
consistent with few 
10$^{5}$$\,$cm$^{-3}$ for this component. Since the fractional
ionization is expected to be of $\leq$10$^{-7}$ in the
ambient cloud \citep{gue82,cas98}, the high--\textit{J} H$^{13}$CO$^{+}$ and
HN$^{13}$C excitation in this component is completely dominated by 
H$_2$ collisions. 

In contrast with the quiescent gas, the precursor component shows an  
over excitation in H$^{13}$CO$^{+}$. The
line ratios and H$_2$ densities estimated for this ion
in the precursor gas, are a factor of 10 larger 
than those for SiO, HN$^{13}$C and H$^{13}$CN. 
In section$\,$4, we have shown that an electron density 
enhancement by a factor of $\sim$500 in this component could
explain the over excitation in H$^{13}$CO$^{+}$. 
%The origin of this electron density enhancement must be related 
%to the shock precursor component. 
Modeling of C--shocks shows that the UV fluorescence
radiation generated by the collisional excitation of H$_2$ in the
magnetic precursor rapidly enhances the ion and electron densities 
in this region \citep[t$\leq$100$\,$yrs from the
inception of the C--shock;][]{flo96,flo03}.
%Thus, the precursor could produce the selective 
%excitation observed in H$^{13}$CO$^{+}$. 
One may think that the probability to detect this enhancement toward molecular
outflows for such short time--scales is negligible. However, L1448--mm 
is a very young outflow (t$_{dyn}$$\sim$1000$\,$yrs), and the
dynamical time--scales derived from the proper motions of the SiO bullets 
\citep[$\sim$90$\,$yrs;][]{gir01} are consistent with the possibility
of detecting the shock precursor as predicted
by C--shock models. In fact, the
%Since H$^{13}$CO$^{+}$ is
%enhanced in the shock precursor toward L1448--mm, and
%HN$^{13}$C and H$^{13}$CN are progressively enhanced in this 
%component from L1448--mm (0,0) to (0,--20). Since 
%the ions and neutrals have slipped to 
more redshifted velocities toward L1448--mm (0,--10) and (0,--20) are
consistent with the observation of different evolutionary stages of C--shocks 
in the different positions in the outflow as predicted by models.

We can use our results to constrain the electron density
enhancement produced by the precursor. 
For optically thin emission and low temperatures, 
the derived fractional ionization is
$\sim$10$^{-3}$. Flower et al. (1996) and Flower \& Pineau des For\^ets
(2003) predicted that the ionization fraction is increased to
$\sim$10$^{-5}$ in the precursor stage. 
Our estimate of the fractional ionization clearly
exceeds these results by a factor of 100, and even exceeds 
the cosmic abundance of atomic carbon \citep[the main repository of
positive charge in dark clouds;
$\chi$(C)$\sim$2--3$\times$10$^{-4}$;][]{car96} by a factor of
$\sim$5. However, if we increase the kinetic temperature \citep[ions
are expected to be rapidly heated by the precursor;][]{dra80} and
consider optically thick emission, 
the ionization fraction can be decreased to
$\sim$5$\times$10$^{-5}$. This result is consistent with the
model predictions \citep{flo96,flo03}.

We cannot rule out the possibility that the ion and electron
enhancement in the precursor component is produced by the radiative
precursor of J--shocks \citep{shu79}. Chemical models that include 
illumination by UV photons 
predict the enhancement of molecules such as HCO$^{+}$, HCO and HCN 
for t$\leq$300$\,$yrs \citep{vit99,vit03}. However, 
HCO \citep[a typical PDR tracer;][]{sch88} is not detected in the precursor
component \citep{jim04}, which suggests that the ion and 
electron enhancements are likely due to the magnetic precursor of
C--shocks.      

In summary, the differences in the kinematics and excitation between
the ion and neutral components in the L1448--mm molecular outflow 
are clear indicators of the early interaction 
of C--shocks with the ambient gas. The over excitation in H$^{13}$CO$^{+}$ 
has allowed to measure, for the first time, the electron density 
enhancement in the precursor of a C--shock. The estimated 
fractional ionization in the precursor component is of 
$\sim$5$\times$10$^{-5}$, which implies
an enhancement of the electron densities by a factor of $\geq$500 with
respect to the ambient gas. 

%TABLE 3--2nd version*******************************************************
\begin{deluxetable}{lcccc}
\tabletypesize{\scriptsize}
\tablecaption{Derived parameters of SiO, ion and neutral species toward
  L1448--mm. \label{tbl-1}}
\tablewidth{0pt}
\startdata

& \multicolumn{3}{c}{TABLE 3} & \\ \hline \hline

Molecule &
\multicolumn{2}{c}{H$_2$$\,$density$\,$(cm$^{-3}$)} &
\multicolumn{2}{c}{Column density$\,$(cm$^{-2}$)} \\ \hline
& \multicolumn{4}{c}{\bf{(0,0)}} \\
& {\bf Ambient} & {\bf Precursor} & {\bf Ambient} & {\bf Precursor} \\

H$^{13}$CO$^{+}$ & 1.3$\times$10$^5$ & 2.0$\times$10$^6$ &
6.7$\times$10$^{11}$ & 2.2$\times$10$^{12}$ \\ 

SiO & $\ldots$ & 2.7$\times$10$^5$ & $\ldots$ & 1.4$\times$10$^{11}$ \\

HN$^{13}$C & 1.9$\times$10$^5$ & $\leq$3.4$\times$10$^5$ &
1.4$\times$10$^{12}$ & 5.1$\times$10$^{11}$ \\

H$^{13}$CN & $\leq$1.8$\times$10$^6$ & $\leq$1.6$\times$10$^6$ &
2.1$\times$10$^{11}$ & 2.4$\times$10$^{11}$ \\

& \multicolumn{4}{c}{\bf{(0,-10)}} \\

H$^{13}$CO$^{+}$ & $\leq$1.3$\times$10$^5$ & 6.0$\times$10$^6$ & 
4.2$\times$10$^{11}$ & 1.8$\times$10$^{12}$ \\

SiO & $\ldots$ & 4.0$\times$10$^5$ & $\ldots$ & 1.3$\times$10$^{11}$ \\

HN$^{13}$C & $\leq$1.9$\times$10$^5$ & $\leq$1.1$\times$10$^5$ & 
1.1$\times$10$^{12}$ & 1.9$\times$10$^{12}$ \\

H$^{13}$CN & $\leq$1.3$\times$10$^6$ & $\leq$1.2$\times$10$^6$ & 
3.0$\times$10$^{11}$ & 3.4$\times$10$^{11}$ \\

& \multicolumn{4}{c}{\bf{(0,-20)}} \\

H$^{13}$CO$^{+}$ & $\leq$1.6$\times$10$^5$
& $\leq$5.7$\times$10$^4$ & 3.1$\times$10$^{11}$ & 1.0$\times$10$^{12}$ \\

SiO & $\ldots$ & 4.5$\times$10$^5$ & $\ldots$ & 1.3$\times$10$^{11}$ \\

HN$^{13}$C & $\leq$4.0$\times$10$^5$ &
$\leq$1.3$\times$10$^5$ & 4.5$\times$10$^{11}$ & 1.6$\times$10$^{12}$ \\

H$^{13}$CN & $\leq$2.0$\times$10$^6$ &
$\leq$1.2$\times$10$^6$ & 1.9$\times$10$^{11}$ & 3.2$\times$10$^{11}$ \\ \hline
\enddata
\end{deluxetable}

%***************************************************************************

\acknowledgments

We acknowledge the Spanish MEC for the support provided through projects
number AYA2002--10113--E, AYA2003--02785--E, ESP2004--00665, and
``Comunidad de Madrid'' Government under PRICIT project
S--0505$/$ESP--0277 (ASTROCAM). This work has benefited from research 
funding from the European Community's Sixth Framework Programme.


\begin{thebibliography}{}

\bibitem[Bhattacharyya et al.(1981)]{bha81}
Bhattacharyya, S. S., Bhattacharyya, B., \& Narayan, M. V. 1981, \apj, 247, 936

\bibitem[Cardelli et al.(1996)]{car96}
Cardelli, J. A., Meyer, D. M., Jura, M., \& Savage, B. D. 1996, \apj,
467, 334

\bibitem[Caselli et al.(1998)]{cas98}
Caselli, P., Walmsley, C. M., Terzieva, R., \& Herbst, E. 1998, \apj,
499, 234

\bibitem[Chu \& Dalgarno(1974)]{chu74}
Chu, S.--I., \& Dalgarno, A. 1974, Phys. Rev., A10, 788 

\bibitem[Curiel et al.(1999)]{cur99}	
Curiel, S., Torrelles, J. M., Rodr\'{\i}guez, L. F., G\'omez, J. F., \&
Anglada, G. 1999, \apj, 527, 310

\bibitem[Draine(1980)]{dra80}
Draine, B. T. 1980, \apj, 241, 1021

\bibitem[Faure \& Tennyson(2001)]{fau01}
Faure, A., \& Tennyson, J. 2001, \mnras, 325, 443

\bibitem[Faure \& Tennyson(2006)]{fau06}
Faure, A., \& Tennyson, J. 2006, in preparation
 
\bibitem[Flower(1999)]{flo99}
Flower, D. R. 1999, \mnras, 305, 651

\bibitem[Flower et al.(1996)]{flo96}
Flower, D. R., Pineau des For\^ets, G., Field, D., \& May, P. W. 1996,
\mnras, 280, 447

\bibitem[Flower \& Pineau des For\^ets(2003)]{flo03}
Flower, D. R., Pineau des For\^ets, G. 2003, \mnras, 343, 390

\bibitem[Girart \& Acord(2001)]{gir01}
Girart, J. M., \& Acord, J. M. P. 2001, \apj, 552, L63

\bibitem[Green \& Thaddeus(1974)]{gre74}
Green, S., \& Thaddeus, P. 1974, \apj, 191, 653

\bibitem[Gu\'elin et al.(1982)]{gue82}
Gu\'elin, M., Langer, W. D., \& Wilson R. W. 1982, \aap, 107, 107

%\bibitem[Haese \& Woods(1979)]{hae79}
%Haese, N. N., \& Woods, R. C. 1979, Chem. Phys. Letters, 61, 396

\bibitem[Jim\'enez--Serra et al.(2004)]{jim04}
Jim\'enez--Serra, I., Mart\'{\i}n--Pintado, J.,
Rodr\'{\i}guez--Franco, A., \& Marcelino, N. 2004, \apj, 603, L49

\bibitem[Langer(1985)]{lan85}
Langer, W. D. 1985, in Protostars and Planets II, ed. D. C. Black \&
M. S. Matthews (Tucson: Univ. Arizona Press), 650

%\bibitem[Lefloch et al. (1998)]{lef98} 
%Lefloch, B., Castets, A., Cernicharo, J., \& Loinard, L. 1998, 
%\apjl, 504, L109

\bibitem[Saha et al.(1981)]{saha81}
Saha, S., Ray, S., Bhattacharyya, B., \& Barua, A. K. 1981, 
Phys. Rev., A23, 2926 

\bibitem[Schenewerk et al.(1988)]{sch88}
Schenewerk, M. S., Snyder, L. E., Hollis, J. M., Jewell, P. R., \&
Ziurys, L. M. 1988, \apj, 328, 785

\bibitem[Shull \& McKee(1979)]{shu79}
Shull, J. M., \& McKee, C. F. 1979, \apj, 227, 131

%\bibitem[Stogryn \& Stogryn(1966)]{sto66}
%Stogryn, D. E., \& Stogryn, A. P. 1966, Mol. Phys., 11, 371

\bibitem[Turner et al.(1992)]{tur92}
Turner, B. E., Chan, K.--W., Green, S., \& Lubowich, D. A. 1992, \apj,
399, 114

\bibitem[Viti \& Williams(1999)]{vit99}
Viti, S., \& Williams, D. A. 1999, \mnras, 310, 517

\bibitem[Viti et al.(2003)]{vit03}
Viti, S., Girart, J. M., Garrod, R., Williams, D. A., \& Estalella, R.
2003, \aap, 399, 187

\end{thebibliography}
\end{document}